\documentclass[aps,prb,amsmath,amssymb,reprint, groupedaddress,showpacs]{revtex4-2}
\usepackage{color, graphicx}     
\usepackage{dcolumn}     
\usepackage{bm}                  
\usepackage[colorlinks=true,urlcolor=blue,linkcolor=blue,citecolor=blue]{hyperref}%
\usepackage[figure,figure*]{hypcap}
\usepackage{amssymb}
\usepackage{latexsym}
\usepackage{amsfonts}
\usepackage{amsmath}
\usepackage{multirow}
\usepackage[dvipsnames]{xcolor}
\usepackage{times}
\UseRawInputEncoding
\begin{document}

	\makeatletter
	\newcommand*\bigcdot{\mathpalette\bigcdot@{.3}}
	\newcommand*\bigcdot@[2]{\mathbin{\vcenter{\hbox{\scalebox{#2}{$\m@th#1\bullet$}}}}}
	\makeatother

	\title
	{{Pseudospin-Induced Asymmetric Field in Non-Hermitian Photonic Crystals with Multiple Topological Transitions}}
	\author{Yuchen Peng}
	\affiliation{Key Laboratory for Micro/Nano Optoelectronic Devices of Ministry of Education $\textit{\&}$ Hunan Provincial Key Laboratory of Low-Dimensional Structural Physics and Devices, School of Physics and Electronics, Hunan University, Changsha 410082, China}%
	\author{Aoqian Shi}
	\affiliation{Key Laboratory for Micro/Nano Optoelectronic Devices of Ministry of Education $\textit{\&}$ Hunan Provincial Key Laboratory of Low-Dimensional Structural Physics and Devices, School of Physics and Electronics, Hunan University, Changsha 410082, China}%
	\author{Peng Peng}
	\affiliation{Key Laboratory for Micro/Nano Optoelectronic Devices of Ministry of Education $\textit{\&}$ Hunan Provincial Key Laboratory of Low-Dimensional Structural Physics and Devices, School of Physics and Electronics, Hunan University, Changsha 410082, China}%
	\author{Jianjun Liu}
	\email{Corresponding author: jianjun.liu@hnu.edu.cn}
	\affiliation{Key Laboratory for Micro/Nano Optoelectronic Devices of Ministry of Education $\textit{\&}$ Hunan Provincial Key Laboratory of Low-Dimensional Structural Physics and Devices, School of Physics and Electronics, Hunan University, Changsha 410082, China}

	\date{\today}
	
	\begin{abstract}
We report the discovery of multiple topological phase transitions induced by the photonic quantum spin Hall effect (PQSHE) in a non-Hermitian photonic crystal (PhC). When increasing the magnitude of the non-Hermitian terms, the system undergoes transitions from topological corner states to topological edge states and subsequently to topological bulk states. The angular momentum of the wave function of the out-of-plane electric field excited by the chiral source acts as the pseudospin degree of freedom in the PQSHE. Therefore, we consider the introduction of chiral sources with different circular polarizations into non-Hermitian PhCs and observe the emergence of asymmetric field responses. These results are expected to enable the multi-dimensional manipulation of topological states, offering a new avenue for the detection of chiral sources.
	 
	\end{abstract}
	\maketitle
 \section{Introduction}
Non-Hermitian systems refer to systems with complex Hamiltonian eigenvalues~\cite{RN1,RN2,RN3}. When the parity-time (PT) inversion symmetry is preserved, the non-Hermitian Hamiltonian exhibits a completely real energy spectrum~\cite{RN4}, often leading to distinct effects and advantages compared with Hermitian systems. Recent advancements in the non-Hermitian system theory and its association with exceptional points (EPs) have revolutionized researchers’ understanding of physical systems with complex eigenvalues~\cite{RN5,RN6,RN7,RN8,RN9,RN10,RN11,RN12}. Studies on non-Hermitian physics have gradually broadened from quantum condensed matter systems to classical wave systems, e.g., optical~\cite{ RN13,RN14,RN15,RN16,RN17,RN18,RN19,RN20} and acoustic~\cite{RN21,RN22,RN23,RN24,RN25,RN26,RN27} systems, demonstrating the possibility of manipulating various physical phenomena.

On the other hand, the discovery of higher-order topological states in photonic crystals (PhCs) has sparked considerable attention in topological corner states (TCSs), paving the way for novel developments of related photonic devices~\cite{RN28,RN29,RN30,RN31,RN32}. Topological photonics offers various degrees of freedom (DOFs), providing a novel approach for light manipulation~\cite{RN33,RN34}. For instance, the pseudospin DOF in the photonic quantum spin Hall effect (PQSHE) can induce the generation of a source with different circular polarizations that excites TCSs with decreasing strength in the counterclockwise or clockwise direction ~\cite{RN35,RN36}. When introducing non-Hermitian terms into topological systems, these terms provide a versatile platform for manipulating the physical properties of TCSs~\cite{RN23,RN37,RN38}. However, previous studies have mainly concentrated on investigating the impact of non-Hermitian terms on the band structures and eigenfields~\cite{RN4,RN7,RN8,RN17,RN18,RN23,RN37,RN39,RN40,RN41,RN42,RN43}. We aim to go beyond this and explore the transmission effects of electromagnetic waves excited by different types of sources, which in turn allows us to observe how the pseudospin DOF can manipulate the transmission in non-Hermitian classical wave systems. The coupling between non-Hermitian terms and the pseudospin DOF in manipulating the characteristics of TCSs may yield intriguing physical effects and novel optical phenomena that merit deeper investigation.
	
		\begin{figure*} 
		\centerline{\includegraphics[scale=.32]{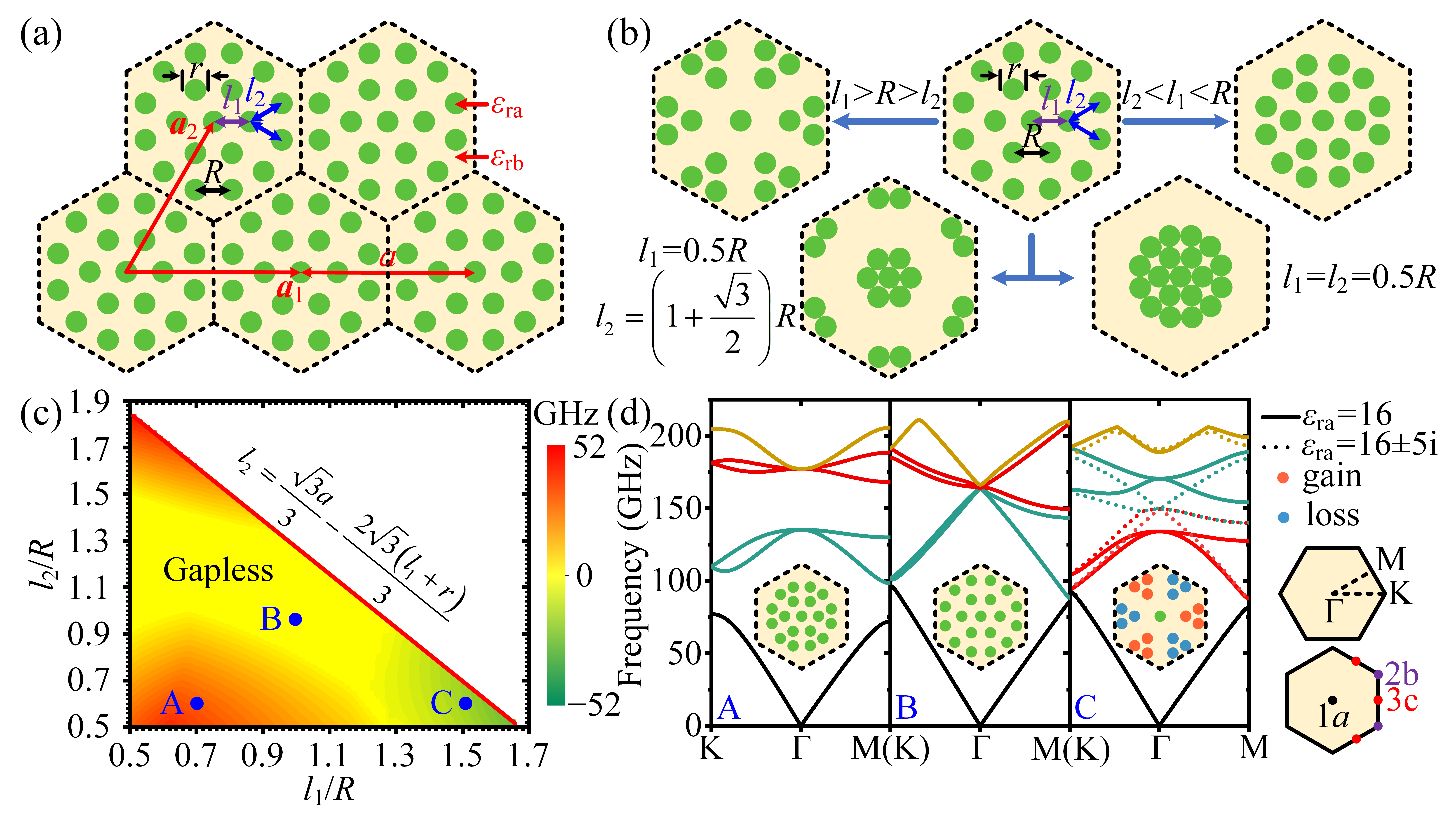}}
		\caption{\label{Fig.1}(a) Schematic of S-T PhCs. (b) Several different types of basic structural units of S-T PhCs and their expanded and shrunken limit states. (c) Schematic of the topological phase transition (green (red): topological nontrivial (trivial) PhCs; yellow: no PBG near 150 GHz with Dirac points in the PBG). (d) Three representative points in Fig. 1(c) corresponding to the band structure of three different types of S-T PhCs, where C corresponds to the introduction of gain (red, type-I scatterers) and loss (blue, type-II scatterers) terms in expanded S-T PhCs, and central scatterers without gain and loss terms (green, type-III scatterers) result in band degeneracy once more. The insets in the bottom right corner show the first Brillouin zone and high-symmetry points in the reciprocal space (top) and Wyckoff locations in real space (bottom).}
	\end{figure*}
	
Here, we report the discovery of multiple topological transitions in non-Hermitian Stampfli-Triangle (S-T) PhCs and the asymmetric field induced by the pseudospin DOF in the PQSHE. S-T PhCs are composed of the basic structural units of Stampfli-type photonic quasicrystals arranged in a triangular lattice, enabling the realization of the PQSHE~\cite{RN36,RN44}. We first present the variation patterns of photonic bandgaps (PBGs) and topological phase diagrams to identify the topological S-T PhC induced by the PQSHE. In our investigation of the topological states within this structure, we discover a phenomenon similar to the slow-light effect in non-Hermitian topological PhCs. To correctly describe the physical properties of non-Hermitian systems, we develop a three-dimensional complex projected band structure that integrates both the imaginary and real parts of the frequency. Subsequently, we investigate the impact of non-Hermitian systems on the higher-order topological states induced by the PQSHE. Specifically, we explore how these systems influence the emergence of TCSs excited by chiral sources with circular polarization. 
	
\section{Model and theory}
	The structure of S-T PhCs is shown in Fig.~\ref{Fig.1}(a); the lattice constant is $a=$ 1 \text{mm}, the lattice basic vectors are $\boldsymbol {a_\text{1}}= (\text{1,0})$ and $\boldsymbol {a_\text{2}}=(a/\text{2},\sqrt{\text{3}}a/\text{2})$, and the distance between adjacent scatterers is  $R$ (in this case, $R = l_\text{1} = l_\text{2}$). In this PhC structure, apart from the previously mentioned approach of changing the size of the scatterers to realize the PQSHE \cite{RN36,RN44}, such a topological state can also be generated by adjusting the spacing between the scatterers within the basic structural unit. The procedure involves the following steps: 1) Dividing the 18 scatterers, excluding the central scatterer, into six groups, each comprising three scatterers. Subsequently, shrinking and expanding the scatterers, i.e., adjusting the size of $l_\text{2}$. 2) Adjusting the spacing between the six groups of scatterers and the central scatterer, i.e., adjusting the size of $l_\text{1}$. Therefore, two different basic structural units can be obtained, resulting in two extreme cases, as shown in Fig.~\ref{Fig.1}(b). Specifically, when $l_\text{1} = l_{\text{1min}} = \text{0.5}a$, the maximum and minimum values of $l_\text{2}$ are $l_{\text{2max}} = (\text{1}+\sqrt{\text{3}}/\text{2})R $ and $ l_{\text{2min}} = \text{0.5}a$, respectively. We also provide the topological phase transition laws for different parameters, as shown in Fig.~\ref{Fig.1}(c). The PBG shown in Fig.~\ref{Fig.1}(c) is obtained as $\Delta f=f_\text{d}-f_\text{p}$, where $f_\text{d}$ corresponds to the frequency of the d band, and $f_\text{p}$ corresponds to the frequency of the p band. Therefore, when the PhC is in a topological trivial state, $\Delta f > $ 0, while when it is in a topological nontrivial state, $\Delta f < $ 0. Taking into consideration the size of the common PBG, we consider PhCs A ($l_\text{1} = \text{0.7}R$, $l_\text{2} = \text{0.6}R$) and C ($l_\text{1} = \text{1.52}R$, $l_\text{2} = \text{0.6}R$) in Fig.~\ref{Fig.1}(c) as the subjects of our investigation. 	For a structure that satisfies $C_n$ symmetry, its band topology can be judged by calculating the topological index, which is extracted from the representation of symmetry at the high symmetry points (HSPs) $\Pi_p^{(n)}$ of all bands below the PBG, where $p=\text{1,2,3,}...\text{,}n$ is the eigenvalue sequence number, the rotation eigenvalues corresponding to each item are $\Pi _p=\mathrm{e}^{2\pi\mathrm{i}(p-\text{1})/n}$~\cite{RN50,RN51}. For the S-T PhC proposed in this paper, whether expanded or shrunken, it always maintains $C_\text{6}$ symmetry, so there are three HSPs: $\Gamma$, K, and M point, within the Brillouin zone. These points are isomorphic to the point groups $C_\text{6}$, $C_\text{3}$, and $C_\text{2}$, respectively. In other words, the irreducible representation corresponding to the HSP can be judged from the characteristic table of the relevant point group. According to previous theories~\cite{RN52,RN53}, the topological phase transition is often accompanied by band inversion, and the number of bands with specific eigenvalues at each HSP remains consistent, which means that before and after the band inversion, there must be a change in the wave function at the HSP below the PBG. This change will lead to a shift in the position of the Wannier center that determines the topological properties of systems. It can be defined whether it is a topological state by comparing the difference between the wave functions of different HSPs. Therefore, the rotation eigenvalues of HSPs $\Pi_p^{(n)}$ are compared with the reference point $\Gamma_p^{(n)}$ to define a new topological invariant with symmetry-indicator~\cite{RN54,RN55}.
	\begin{equation}
		\label{eq.1}
		[\Pi_p^{(n)}]=\#\Pi_p^{(n)}-\#\Gamma_p^{(n)},
	\end{equation}
	where $\#\Pi_p^{(n)}$ represents the number of bands with $C_n$ symmetry rotation eigenvalues at the HSP $\Pi=\Gamma$, K, M below the PBG, and then it can be determined whether a structure that satisfies $C_\text{6}$ symmetry possesses a topological state through the topological index $\chi^{(6)}=([\text{M}_1^{(2)}],[\text{K}_1^{(3)}])$ ~\cite{RN54,RN55}. If there is a non-zero item in the topological index, the S-T PhC is in the topological nontrivial state, otherwise, it is in the topological trivial state. On the other hand, to judge whether there is a TCS in the S-T PhC, it is necessary to calculate whether the quadrupole mode $Q_c$ is non-zero:
	\begin{equation}
		\label{eq.2}
		Q_c=\frac{1}{4}[\text{M}_1^{(2)}]+\frac{1}{6}[\text{K}_1^{(3)}].
	\end{equation}
	
	The rotation eigenvalue of a PhC can be judged by the phase distribution and the specific analysis of the topological states for PhCs A and C can be seen in the section I of Supplemental Material~\cite{RN49}. As shown in Fig.~\ref{Fig.1}(d), it is evident that the previously open PBG closes again when $\gamma = $ 5. This suggests the potential existence of a topological phase transition within this process, which warrants further in-depth exploration.
	
\begin{figure} 
	\centerline{\includegraphics[scale=.35]{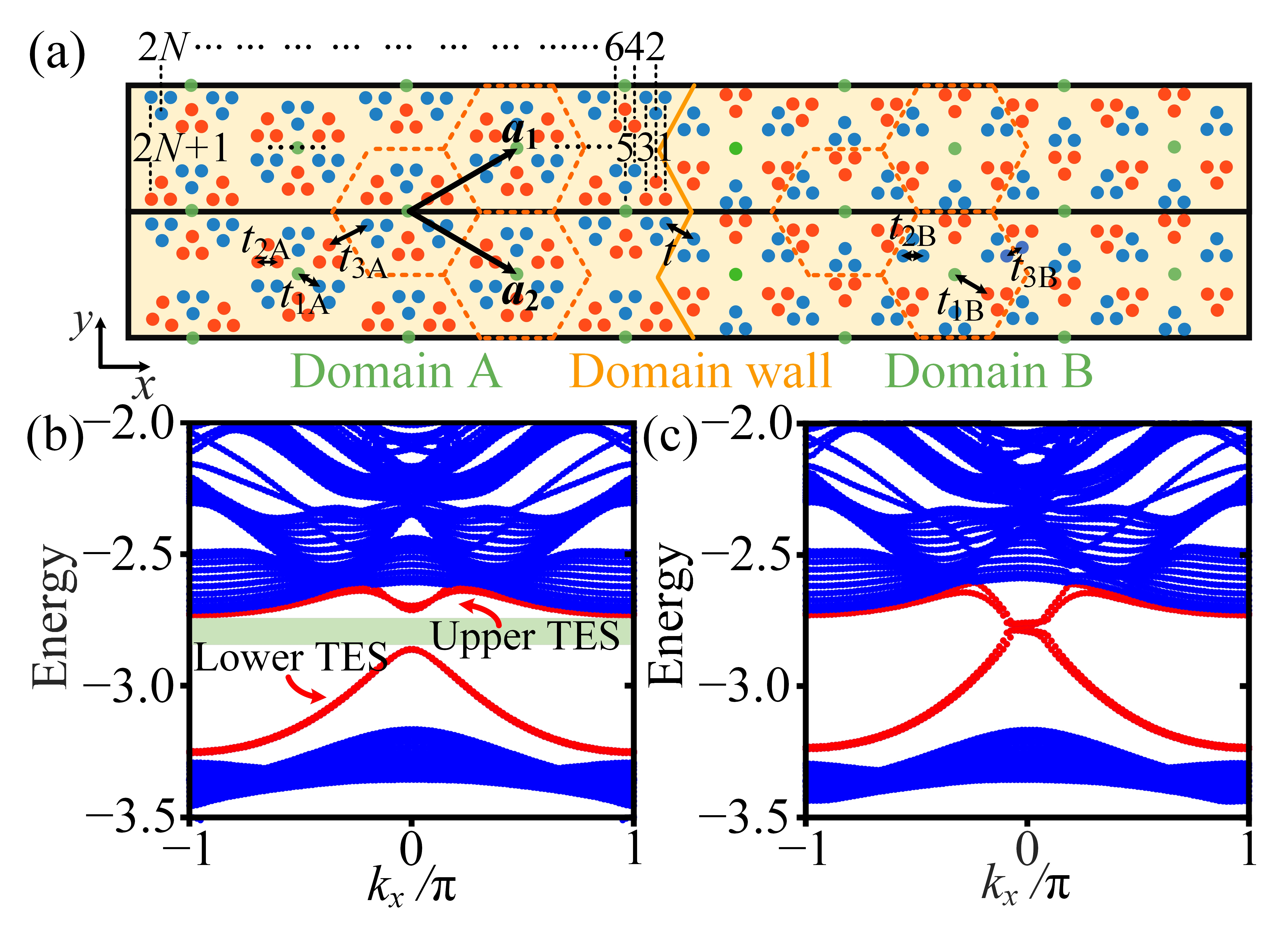}}
	\caption{\label{Fig.2}(a) Schematic of the strip supercell of S-T PhCs. Dispersion of the TESs for (b) $\gamma =$ 0 and (c) $\gamma =$ 0.16.}
\end{figure}

	To verify the topological phase transition, topological edge states (TESs) should be found in the PBG. In Fig.~\ref{Fig.2}(a), the supercell is divided into two regions: one with expanded S-T PhCs, denoted as Domain A, and the other with shrunken S-T PhCs, denoted as Domain B. A domain wall is formed at the interface between these two domains. Subsequently, the physical model of the supercell shown in Fig.~\ref{Fig.2}(a) is derived to study the impact of non-Hermitian terms on the topological states. 

	According to the second quantization formalism, the tight-binding Hamiltonian can be expressed as follows~\cite{RN39}:
	\begin{equation}
		\label{eq.3}
		H=-t\sum_{\substack{<i,j>}} c_i^\dagger c_j - \sum_{\substack{s = \text{A,B}}} \sum_{\substack{i}} m_{s,i} c_{s,i}^\dagger c_{s,i} + h.c.\;\text{,}\;
	\end{equation}
	where $t$ represents the nearest-neighbor coupling term,  $c_i^\dagger$ represents the creation operator, $c_j$  represents the annihilation operator, and $m_{s,i}$ represents the $i$-th ($i =$ I, II, III)-type scatterers in Domain $s$ ($s =$ A, B) with gain and loss terms. After substituting Eq.~\eqref{eq.3} into the Schr{\"o}dinger equation, the wave function of each supercell can be derived as~\cite{RN39}:
	\begin{equation}
		\label{eq.4}
	    | \psi (k) \rangle =\sum_{\substack{s,i,n}} \psi_{s,i} (n)  c_{s,i,n}^ \dagger(k) | G \rangle \;\text{,}\;
	\end{equation}
	where $ | G \rangle$ represents the ground state of the Hamiltonian, $n$ represents the label of the scatterers in the supercell, and $\psi_{s,i} (n)$ represents the wave function of the $n$-th scatterer at the $i$-th lattice point in Domain $s$ in Fig.~\ref{Fig.2}(a). For instance, $\psi_{\text{A},I} (n)$ corresponds to the wave function of the gain-introduced red scatterer labeled $n$ in Domain A. Substituting Eq.~\eqref{eq.4} into the Schr{\"o}dinger equation yields the corresponding equations of motion for each part of the tight-binding model. For Domain $s$ ($s=$ A, B):
	\begin{equation}
		\label{eq.5}
		\begin{split}
		&\epsilon \psi _{s \text{,I}}(n)  = - \psi _{s \text{,I}}(n+\text{1})-\psi _{s \text{,I}}(n+\text{2})-\psi _{s \text{,II}}(n-\text{2})\\
		&-\mathrm{e}^{- \mathrm{i} \Pi_\text{1}k}\psi _{s \text{,II}}(n) - \Lambda_\text{1} \psi _{s \text{,III}}(n-\text{2})-\gamma \psi _{s \text{,I}}(n) \;\text{,}\;\\
		&\Pi_\text{1}=\left\{
		\begin{aligned}
			\text{0} & \text{, } n=\text{6}j+\text{4} \\
			\text{1} & \text{, } n =\text{6}j+\text{1}\\
		\end{aligned}
		\right.\text{, } j=\text{0,1,2,}...\text{,}N\text{,}\\
		&\Lambda_\text{1}=\left\{
		\begin{aligned}
			\text{0} & \text{, } n=\text{9}j+\text{1},\text{9}j+\text{4} \\
			\text{1} & \text{, } n =\text{9}j+\text{7}\\
		\end{aligned}
		\right.\text{, } j=\text{0,1,2,}...\text{,}N\text{,}
	\end{split}
\end{equation}

		\begin{equation}
			\label{eq.6}
		\begin{split}
			&\epsilon \psi _{s \text{,I}}(n)  = - \psi _{s \text{,I}}(n-\text{1})-\psi _{s \text{,I}}(n+\text{1})-\psi _{s \text{,II}}(n+\text{2})\\
			&-\psi _{s \text{,II}}(n-\text{2}) - \Lambda_\text{2} \psi _{s \text{,III}}(n)-\gamma \psi _{s \text{,I}}(n) \;\text{,}\;\\
			&\Lambda_\text{2}=\left\{
			\begin{aligned}
				\text{0} & \text{, } n=\text{9}j+\text{2, }\text{9}j+\text{8}\\
				\text{1} & \text{, } n =\text{9}j+\text{5}\\
			\end{aligned}
			\right.\text{, } j=\text{0,1,2,}...\text{,}N\text{,}
		\end{split}
	\end{equation}
	
		\begin{equation}
			\label{eq.7}
		\begin{split}
			&\epsilon \psi _{s \text{,I}}(n)  = - \psi _{s \text{,I}}(n-\text{1})-\psi _{s \text{,I}}(n-\text{2})-\psi _{s \text{,II}}(n+\text{2})\\
			&-\mathrm{e}^{- \mathrm{i} \Pi_\text{2}k}\psi _{s \text{,II}}(n) - \Lambda_\text{3} \psi _{s \text{,III}}(n+\text{2})-\gamma \psi _{s \text{,I}}(n) \;\text{,}\;\\
			&\Pi_\text{2}=\left\{
			\begin{aligned}
				\text{0} & \text{, } n=\text{6}j+\text{6} \\
				\text{1} & \text{, }  n=\text{6}j+\text{3}\\
			\end{aligned}
			\right.\text{, } j=\text{0,1,2,}...\text{,}N\text{,}\\
			&\Lambda_\text{3}=\left\{
			\begin{aligned}
				\text{0} & \text{, } n=\text{9}j+\text{2}\text{, } \text{9}j+\text{9} \\
				\text{1} & \text{, } n =\text{9}j+\text{3}\\
			\end{aligned}
			\right.\text{, } j=\text{0,1,2,}...\text{,}N\text{,}
		\end{split}
	\end{equation}
	Here, $\Pi$ and $\Lambda$ are symbolic indicators to determine whether a transition occurs. Each symbol in Eqs.~\eqref{eq.5}-\eqref{eq.7} is exclusively determined by the corresponding $n$,  which reveals the presence of intercell couplings among the scatterers as well as the coupling strength between the central scatterer and the other scatterers. These symbols can also be used to distinguish the wave function $\psi_{s,i} (n)$ of each scatterer with distinct labels, and the number of scatterers in the supercell should be theoretically infinite, i.e., $N\rightarrow \infty$. According to the equations of motion provided above, the Hamiltonian based on the tight-binding model can also be written in a matrix form. The basic periodic unit $A_{i,j}$ ($i,j =\text{1, 2, } ...\text{,  38}$) within Domain A is represented as a 38$\times$38 matrix in real space. The specific value of each element of the matrix corresponds to the coupling of the scatterer with its nearest-neighbor scatterer. Details regarding the nearest-neighbor scatterer for each scatterer can be obtained from Eqs.~\eqref{eq.5}-\eqref{eq.7}. Similarly, within Domain B, corresponding basic periodic units $B_{i,j}$ ($i,j =\text{1, 2, } ...\text{,  38}$) also exist. A domain wall is formed at the interface between Domains A and B, which is characterized by the Hamiltonian $C_{i,j}$ ($i,j =\text{1, 2, } ...\text{,  38}$). Except for $C_{35,2} = C_{36,3} = C_{37,1} =C_{35,2} = C_{38,4} = (t_4+t_4')/2$, the other matrix elements are all equal to 0. Moreover, the Hamiltonian $D_{i,j}$ ($i,j =\text{1, 2, } ...\text{,  38}$) between each supercell needs to be considered. Therefore, the Hamiltonian for the entire Domain A/B is written as: 
	\begin{equation}
		\label{eq.8}
		H_{A(\text{or } B)}= I \otimes A_{i,j} (\text{or } B_{i,j})+
		\begin{pmatrix}
			O & C_{i,j} & O \\
			C_{i,j}^\dagger & O & C_{i,j} \\
			O & C_{i,j}^\dagger & O
		\end{pmatrix}\text{,}
		\end{equation}
where $I$ is the 38$\times$38 unit matrix, and then the Hamiltonian of the supercell can be obtained as:
	\begin{equation}
		\label{eq.9}
	H= 
	\begin{pmatrix}
		H_A &D_{i,j} +D_{i,j}^\dagger & O \\
		O & H_B & D_{i,j} +D_{i,j}^\dagger \\
		O & O & H_A
	\end{pmatrix}.
\end{equation}

By solving for the eigenvalues of the Hamiltonian, when $\gamma =$ 0, it is found that the TES is not entirely gapless, as shown in Fig.~\ref{Fig.2}(b). However, when $\gamma =$ 0.16, the TESs become degenerate. At this point, the TES changes from a gapped structure to a gapless structure, as shown in Fig.~\ref{Fig.2}(c). The Hamiltonian of the S-T lattice structure derived from the tight-binding model is consistent with the physical properties of the PhC. The difference is that the topological nature of the Hamiltonian is generated by adjusting the transition value, whereas the topological nature of the PhC is generated by adjusting the coupling coefficient through adjustments of the distance between the scatterers. Therefore, based on the Hamiltonian, it is possible to predict the dispersion relationship of the TES when introducing a staggered distribution of the conjugate imaginary part into the relative permittivity of the S-T PhCs, as shown in Fig.~\ref{Fig.2}(a). 

\section{Results and discussions}
\begin{figure} 
	\centerline{\includegraphics[scale=.36]{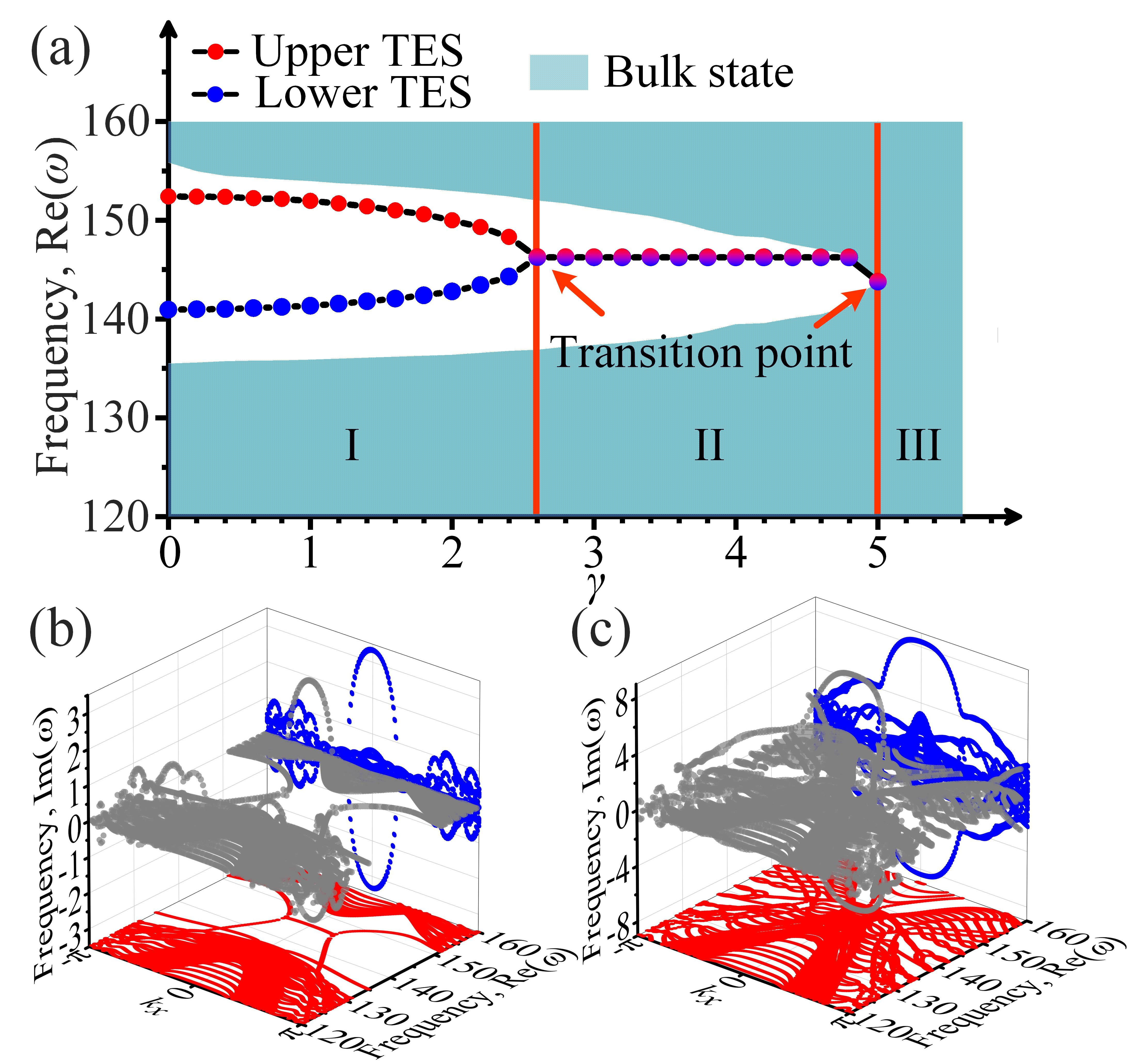}}
	\caption{\label{Fig.3}(a) Schematic illustrating the existence of multiple topological phase transitions. (b) Projected bands corresponding to the phase transition point between regions I and II. (c) Projected bands corresponding to the phase transition point between regions II and III.}
\end{figure}

In this paper, using the finite element method, we determine the impact of the magnitude of the imaginary part of the relative permittivity on the projected band frequencies of the PhC supercell, as shown in Fig.~\ref{Fig.3}.
The projected band in non-Hermitian PhCs has a complex frequency; thus, we need to analyze the physical properties in terms of real and imaginary parts of the frequency. To describe the topological properties of the S-T PhC, we consider the minimum frequency of the upper TESs and the maximum frequency of the lower TESs in terms of the real part of the frequency to represent the variation in the PBG. These values effectively represent the frequency trends of the TESs as a function of $\gamma$ and are depicted as the red and blue dots in Fig.~\ref{Fig.3}(a). The frequency distribution of the topological bulk states (TBSs) is shown in the shaded blue region in Fig.~\ref{Fig.3}(a). By analyzing the evolution of the projected bands, it becomes evident that as the magnitude of the imaginary part γ increases, the PBGs between the upper and lower TESs become narrower. When $\gamma =$ 2.6, the dispersion curves of the upper and lower TESs become degenerate, giving rise to a gapless structure, as shown in Fig.~\ref{Fig.3}(b). This transformation also signifies that the S-T PhC transitions from a higher-order topological state to a first-order topological state. When $\gamma \in$ [2.6,5], the upper and lower TBSs gradually approach each other and become degenerate at $\gamma =$ 5. At this point, the S-T PhC transitions from a first-order topological state to a perfect conductor characterized solely by the TBSs, as shown in Fig.~\ref{Fig.3}(c). To verify the non-Hermitian nature of the S-T PhC, we need to analyze the imaginary part of the frequency. It is evident that its TESs exhibit characteristics resembling those of an exceptional ring~\cite{RN5,RN45}. The endpoint of the ring is the EP of the system, signifying that the S-T PhC is in a PT-broken-symmetry state. In most works of non-Hermitian physics~\cite{RN17,RN46,RN47}, researchers have separated the imaginary and real components of their complex projected bands, which have yielded incorrect descriptions of the flat bands in non-Hermitian TESs (as shown in Fig.~\ref{Fig.3}(b)); indeed, these flat bands have been mistakenly associated with slow-light effects. We developed a three-dimensional complex projected band structure that integrates both the imaginary and real parts of the frequency and projects them onto two planes. When adopting this method, we can determine whether the flat band in the real part of the frequency is actually a projection of the imaginary part of the frequency.

To further confirm the existence of multiple topological phase transitions, the electromagnetic transmission effect induced by a chiral source (the details can be seen in the section II of Supplemental Material~\cite{RN49}) in the box-shaped waveguide is calculated.
\begin{figure} 	
	\centerline{\includegraphics[scale=.42]{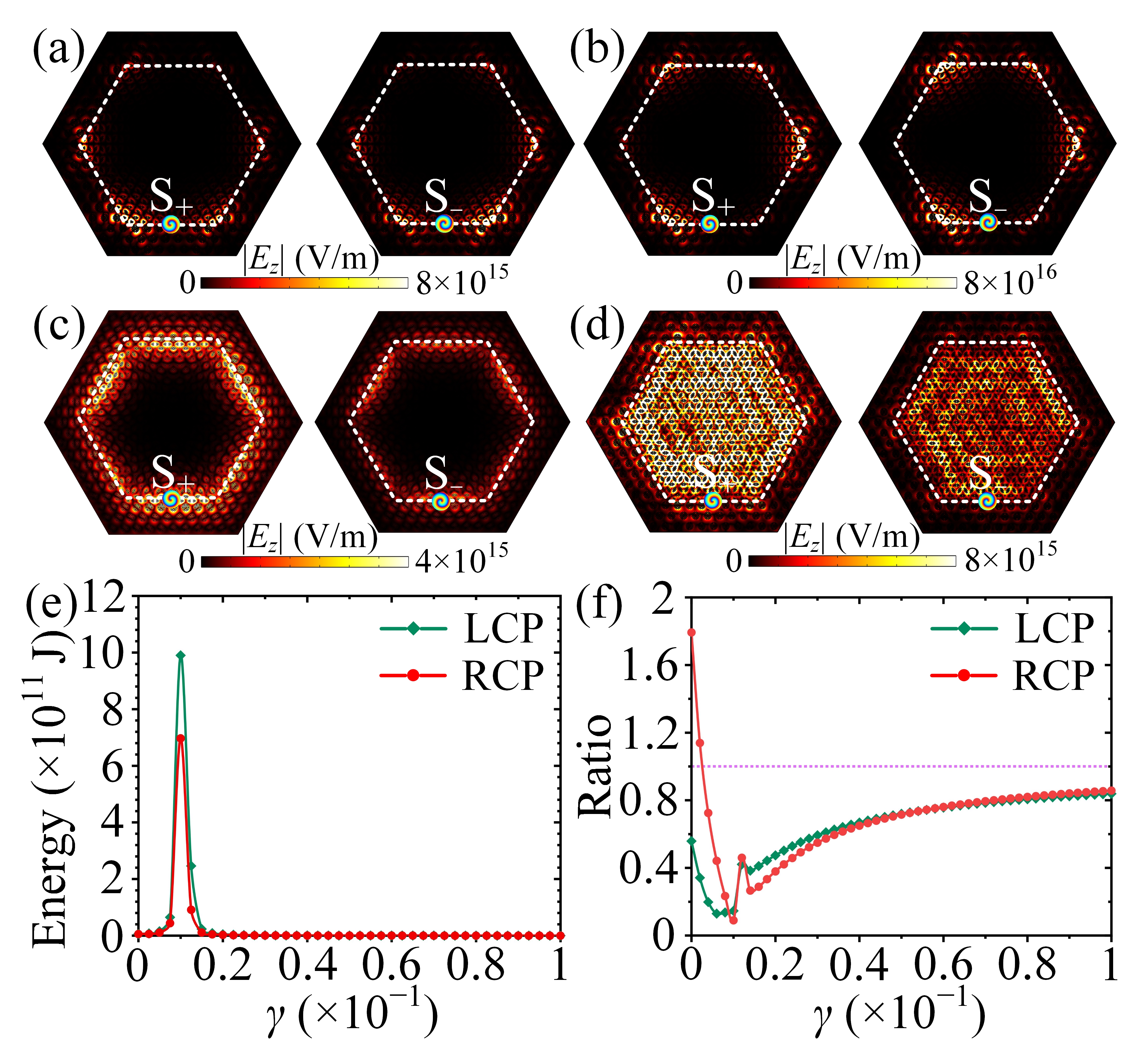}}
	\caption{\label{Fig.4}Fields excited by circularly polarized sources $\text{S}_\pm = H_\text{0}\mathrm{e}^{i\omega t}(\mathrm{e}_x\mp i\mathrm{e}_y)$ ($\text{S}_+$ is the right circularly polarized source (RCPS), $\text{S}_-$ is the left circularly polarized source (LCPS)) at the same frequency (145.21 GHz) with different polarizations. The white hexagonal dotted lines are the boundary between the PhC with topological trivial states and nontrivial states. (a) $\gamma = \text{0}$, (b) $\gamma = \text{0.01}$, (c) $\gamma = \text{2.7}$, (d) $\gamma = \text{5}$; the left (right) images of panels (a)$\--$(d) are the TCSs, TESs, and TBSs excited by $\text{S}_+$ ($\text{S}_-$). (e) System energy as a function of $\gamma$ for $\gamma \in$ [0,0.1]. (f) Ratio of the intensities of the left and right corner fields as a function of $\gamma$ for $\gamma \in$ [0,0.1].}
\end{figure}
As shown in Figs.~\ref{Fig.4}(a) and \ref{Fig.4}(b), the presence of non-Hermitian terms causes the TCSs that initially existed at the six corners of the box-shaped waveguide to concentrate in the two corners nearest to the source at the same frequency. Notably, sources with different circular polarizations exhibit an asymmetric amplification response to the field intensity. Moreover, as shown in Figs.~\ref{Fig.4}(c) and \ref{Fig.4}(d), the source can excite both TESs and TBSs at the same frequency by adjusting the magnitude of the imaginary part of the relative permittivity, demonstrating the existence of multiple topological transitions. Similarly, TESs and TBSs also exhibit the phenomenon of an asymmetric field response induced by the pseudospin. The box-shaped structure surrounded by perfect matching layers is a closed system, so the field intensity of the entire system can be characterized by the energy. Specifically, the energy of the TESs excited by an LCPS (RCPS) is 1.68$\times\text{10}^\text{8}$ J (18.56$\times\text{10}^\text{8}$), and the energy of the TBSs excited by an LCPS (RCPS) is 313.37$\times\text{10}^\text{8}$ J (1148.04$\times\text{10}^\text{8}$ J). It can be seen from Fig.~\ref{Fig.4}(e) that when the imaginary part of the relative permittivity is not introduced, the entire box-shaped structure possesses the same energy. When $\gamma = \text{0.01}$, the overall field intensity in the box-shaped structure has been significantly enhanced. Subsequently, as $\gamma$ increases from 0.01 to 0.1, the field intensity gradually diminishes. Remarkably, for TCSs, the intensity of the field excited by the LCPS consistently exceeds that of the field excited by the RCPS, exhibiting the completely opposite response for TESs and TBSs. Due to the significant disparity in field intensity observed when $\gamma = \text{0.01}$ compared with other values, we calculate the energy variation of the TCSs in an interval of 0.001 centered around $\gamma$ = 0.01. Our investigation reveals that when $\gamma \in$ [0,0.01], the energy of the TCSs gradually increases; on the other hand, when $\gamma \in$ [0.01,0.02], the energy of the TCSs gradually decays (see details in the section III of Supplemental Material~\cite{RN49}). Moreover, as shown in Fig.~\ref{Fig.4}(f), concerning the TCSs, the ratio of the field intensities of the nearest-neighbor left and right corners gradually approaches 1 as $\gamma$ increases. This observation underscores the fact that the non-Hermitian term exerts a stronger control over the TCSs than the pseudospin DOF. 

\section{Conclusions}
In conclusion, we constructed a non-Hermitian S-T PhC and provided a comprehensive theoretical analysis based on the Hamiltonian derived from the tight-binding model and the elementary band representation theory. We described the topological phase transition patterns upon increasing the magnitude of the non-Hermitian parameters, providing evidence for the existence of multiple topological phase transitions. Furthermore, the introduction of chiral sources with different circular polarizations into the PhCs can generate an asymmetric field intensity phenomenon, and we observed a significant energy gain for a specific non-Hermitian parameter. Taking advantage of the fact that TCSs, TESs, and TBSs can be excited at the same frequency, the significant energy gain will overcome the limitations imposed by the coupling among different fields of topological states~\cite{RN17,RN25}. Moreover, it is envisaged that the observed asymmetric response will open a new avenue for identifying the polarization characteristics of chiral sources by measuring the field intensity~\cite{RN48}.

\section*{Acknowledgements}
This work was supported by the National Natural Science Foundation of China (Grants No. 61405058 and No. 62075059), the Natural Science Foundation of Hunan Province (Grants No. 2017JJ2048 and No. 2020JJ4161), and the Scientific Research Foundation of Hunan Provincial Education Department (Grant No. 21A0013).
 
 \bibliography{Manuscript}

	\end{document}